\documentstyle[aps]{revtex}

\begin{document}
\author{Yang lei, Zhu zhengang, Wang yinghai}
\address{({\it Department of physics, Lanzhou University, Gansu 730000, China)}}
\title{{\ Exact solutions of nonlinear equations}}
\maketitle

\begin{abstract}
Based the homogeneous balance method, a general method is suggested to
obtain several kinds of exact solutions for some kinds of nonlinear
equations. The validity and reliability of the method are tested by applying
it to the Bousseneq equation.

\end{abstract}

Nonlinear PDE(partial differential equations) are widely used to describe
complex phenomena in various fields of science, especially in physical
science. A vast variety of methods including B\"{a}cklund transform$^{[1]}$,
Inverse Scattering theory$^{[2]}$ and Hirota's bilinear methods$^{[3]}$ has
been developed to obtain analytic solutions of nonlinear PDE. But in some
cases$^{[4,5]}$, they cannot do very well. So the simple and direct methods
to find analytic solutions of PDE have drawn a lot of interest, for example,
the truncated Painlev\.{e} expansion$^{[6,7]}$, the hyperbolic tangent
function-series method$^{[8,9]}$ and homogeneous balance method$^{[10,11]}$, 
{\it etc}. However, only solitary wave solutions are found in most of those
methods.

Recently, Wang$^{[10,11]}$ showed the homogeneous balance method is powerful
for finding analytic solitary wave solutions of PDE. The essence of the
homogeneous balance method can be presented in this way, the nonlinear PDE\
is 
\begin{equation}
P(u,u_x,u_t,u_{xx},u_{xt},u_{tt},\cdots \cdots )=0.  \label{1}
\end{equation}
If the solution of equation(1) has the form 
\begin{equation}
u=\sum_{i=0}^Na_if^{(i)}(\omega (x,t)),  \label{2}
\end{equation}
where $i$ is an integer, $a_i$ is coefficient, $f(\omega (x,t))$ is a
function of $\omega (x,t)$ only, $\omega (x,t)$ is a function of $x$ and $t$%
, $(i)$ represents derivative index. The nonlinear terms and the highest
order partial derivative terms ought to be partially balanced according to
the assumption of homogeneous balance, so $N$ is obtained. Assuming the form
of $f(\omega (x,t))$ is

\begin{equation}
f=b\ln (1+e^{\alpha x+\beta t+\gamma }).  \label{3}
\end{equation}
Putting (2) and (3) into (1), then deciding coefficients $a_i,b,\alpha
,\beta ,\gamma $, thus the solitary ware solutions of a given nonlinear PDE
are obtained.

This paper noted that the original equation became a set of equations by
using the homogeneous balance method, and we attempt to get other kinds of
exact solutions except solitary wave solution by solving the set of
equations. When $N$ is determined by balancing the nonlinear terms and the
highest order partial derivative terms, the form of $f(\omega (x,t))$ and
the relation of $f^{(i)}\cdot f^{(j)}$ and $f^{(i+j)}$ can be derived. Then
substitute the form of $f(\omega (x,t))$ and the relation of $f^{(i)}\cdot
f^{(j)}$ and $f^{(i+j)}$ into equation (1), we get

\begin{equation}
F(f^{\prime },f^{\prime \prime },\ldots \omega _x,\omega _{xx},\ldots \omega
_t,\omega _{tt},\ldots \omega _{xt},\ldots )=0,  \label{4}
\end{equation}
where $F$ is a function of $f^{\prime },f^{\prime \prime },\ldots \omega
_x,\omega _{xx},\ldots \omega _t,\omega _{tt},\ldots \omega _{xt},\ldots $.
Obviously, $F$ is a linear polynomial of $f^{\prime },f^{\prime \prime
},\ldots $. Setting the coefficients of $f^{\prime },f^{\prime \prime
},\ldots $ to zero yields a set of partial differential equations for $%
\omega (x,t)$. Take note of the term including $f^{\prime }$ in formula (4),
its expression must be $f^{^{\prime }}\omega _{x_1,x_2,\ldots x_p,t_1\ldots
t_q}$. So the coefficient of $f^{^{\prime }}$ in formula (4) is a linear
polynomial $\sum\limits_{i=1}^ka_i\omega _{x_1,x_2,\ldots x_{pi},t_1\ldots
t_{qi}}$, then the set of the partial differential equations has an
important characteristic, namely the equation from the coefficients of $%
f^{\prime }$ is a linear partial differential equations of $\omega (x,t)$.
Other equations can be regarded as limited conditions to the linear
equation, thus solving the set of equations becomes solving the linear PDE
with some limited conditions.

In this paper we choose Bousseneq equation$^{[12]}$

\begin{equation}
u_{tt}-u_{xx}-a(u^2)_{xx}+bu_{xxxx}=0,  \label{5}
\end{equation}
to illustrate our method, where $a,b$ are real constants. The bousseneq
equation( the wave motion propagate in both directions) describes
one-dimensional weakly nonlinear dispersive water waves and can be derived
from Toda Lattice, a suitable approximation enables the KdV equation( the
wave motion is restricted to be one direction) to be derived from this
equation.

Supposing the solution is of the form 
\begin{equation}
u=\sum_{i=0}^Nf^{(i)}(\omega (x,t)),  \label{6}
\end{equation}
substituting (6) into (5), and balancing the nonlinear term $a(u^2)_{xx}$
and the linear term $bu_{xxxx}$, we can get $N=2$,

\begin{equation}
u=f^{^{\prime \prime }}\omega _x^2+f^{^{\prime }}\omega _{xx}.  \label{7}
\end{equation}
Substituting (7) into (5), and collecting all homogeneous terms in partial
dericatives of $\omega (x,t)$, we have

\begin{equation}
\begin{array}{c}
\lbrack bf^{(6)}-2af^{^{\prime \prime \prime }}f^{^{\prime \prime \prime
}}-2af^{^{\prime \prime }}f^{(4)}]\omega _x^6+[15bf^{(5)}-24af^{^{\prime
\prime }}f^{^{\prime \prime \prime }}-2af^{^{\prime }}f^{(4)}]\omega
_x^4\omega _{xx}+ \\ 
\lbrack f^{(4)}\omega _t^2\omega _x^2-f^{(4)}\omega
_x^4+(45bf^{(4)}-24af^{^{\prime \prime }}f^{^{\prime \prime
}}-12af^{^{\prime }}f^{^{\prime \prime \prime }})\omega _x^2\omega
_{xx}^2+(20bf^{(4)}- \\ 
8af^{^{\prime \prime }}f^{^{\prime \prime }}-4af^{^{\prime }}f^{^{\prime
\prime \prime }})\omega _x^3\omega _{xxx}]+[(\omega _{tt}\omega _x^2+\omega
_t\omega _x\omega _{xt}+\omega _t^2\omega _{xx}-6\omega _x^2\omega
_{xx})f^{^{\prime \prime \prime }}+ \\ 
(15bf^{^{\prime \prime \prime }}-6af^{^{\prime }}f^{^{\prime \prime
}})\omega _{xx}^3+(60bf^{^{\prime \prime \prime }}-20af^{^{\prime
}}f^{^{\prime \prime }})\omega _x\omega _{xx}\omega _{xxx}+(15bf^{^{\prime
\prime \prime }}-2af^{^{\prime }}f^{^{\prime \prime }})\omega _x^2\omega
_{xxxx}]+ \\ 
\lbrack (2\omega _{xt}^2+2\omega _x\omega _{xtt}+\omega _{tt}\omega
_{xx}-3\omega _{xx}^2+2\omega _t\omega _{xxt}-4\omega _x\omega
_{xxx})f^{^{\prime \prime }}+(10bf^{^{\prime \prime }}-2af^{^{\prime
}}f^{^{\prime }})\omega _{xxx}^2+ \\ 
(15bf^{^{\prime \prime }}-2af^{^{\prime }}f^{^{\prime }})\omega _{xx}\omega
_{xxxx}+6bf^{^{\prime \prime }}\omega _x\omega _{xxxxx}]+(\omega
_{xxtt}+b\omega _{xxxxxx}-\omega _{xxxx})f^{^{\prime }}=0.
\end{array}
\label{8}
\end{equation}
Setting the coefficient of $\omega _x^6$ in (8) to zero yields an ordinary
differential equation for $f$

\begin{equation}
bf^{(6)}-2af^{^{\prime \prime \prime }}f^{^{\prime \prime \prime
}}-2af^{^{\prime \prime }}f^{(4)}=0.  \label{9}
\end{equation}
Solving (9) we obtain a solution

\begin{equation}
f=-\frac{6b}a\ln \omega ,  \label{10}
\end{equation}
which yields

\begin{equation}
\begin{array}{l}
f^{^{\prime \prime }}f^{^{\prime \prime \prime }}=\frac b{2a}f^{(5)},\qquad
f^{^{\prime }}f^{(4)}=\frac{3b}{2a}f^{(5)},\qquad f^{^{\prime \prime
}}f^{^{\prime \prime }}=\frac baf^{(4)}, \\ 
f^{^{\prime }}f^{^{\prime \prime \prime }}=\frac{2b}af^{(4)},\qquad
f^{^{\prime }}f^{^{\prime \prime }}=\frac{3b}af^{^{\prime \prime \prime
}},\qquad \ \ f^{^{\prime }}f^{^{\prime }}=\frac{6b}af^{^{\prime \prime }}.
\end{array}
\label{11}
\end{equation}
Substituting (11) into (8), formula (8) can be simplified to a linear
polynomial of $f^{\prime },f^{\prime \prime },\ldots $, then setting the
coefficients of $f^{\prime },f^{\prime \prime },\ldots $ to zero yields a
set of partial differential equations for $\omega (x,t)$,

\begin{equation}
\omega _t^2-\omega _x^2-3b\omega _{xx}^2+4b\omega _x\omega _{xxx}=0,
\label{12}
\end{equation}

\begin{equation}
\omega _{tt}\omega _x^2+4\omega _t\omega _x\omega _{xt}+\omega _t^2\omega
_{xx}-6\omega _x^2\omega _{xx}-3b\omega _{xx}^3+9b\omega _x^2\omega
_{xxxx}=0,  \label{13}
\end{equation}

\begin{equation}
2\omega _{xt}^2+2\omega _x\omega _{xtt}+\omega _{tt}\omega _{xx}-3\omega
_{xx}^2+2\omega _t\omega _{xxt}-4\omega _x\omega _{xxx}-2b\omega
_{xxx}^2+3b\omega _{xx}\omega _{xxxx}+6b\omega _x\omega _{xxxxx}=0,
\label{14}
\end{equation}

\begin{equation}
\omega _{xxtt}+b\omega _{xxxxxx}-\omega _{xxxx}=0.  \label{15}
\end{equation}
Where equation (15) from the coefficients of $f^{\prime }$ is a linear PDE.

First, we discuss the travelling wave solution of equation(15). Let $\xi
=x-vt$ and $\omega (x,t)=s(\xi )$, equation (15) becomes an ordinary
differential equation, 
\begin{equation}
bs^{(6)}+(v^2-1)s^{(4)}=0.  \label{16}
\end{equation}
The solution of equation (16) is easy gotten and there exist some different
cases to discuss further.

Case 1, If $\beta =\sqrt{\frac{1-v^2}b}>0$, the solution of equation(15) is 
\begin{equation}
\omega (x,t)=s(\xi )=d_0+d_1(x-vt)+d_2(x-vt)^2+d_3(x-vt)^3+d_4e^{\beta
(x-vt)}+d_5e^{-\beta (x-vt)}  \label{17}
\end{equation}
Substituting formula(17) into limited conditions(12)-(14), a set of algebra
equations are gotten, solving the set of equations, we get $d_1=0$, $d_2=0$, 
$d_3=0$, $d_4=0$ or $d_5=0$. Thus the solutions of equations(12)-(15) are
gotten,

\begin{equation}
\omega (x,t)=d_0+d_4e^{\beta (x-vt)};  \label{18}
\end{equation}

\begin{equation}
\omega (x,t)=d_0+d_5e^{-\beta (x-vt)}.  \label{19}
\end{equation}
Where $d_0$, $d_4$or $d_5$ are arbitrary constants. Substitute the solution
(18) or (19) into formula(6), the exact solitary wave solutions of
equation(5) are obtained,

\begin{equation}
u(x,t)=-\frac{3(1-v^2)}{2a}\cosh ^{-2}(\frac 12(\beta (x-vt)+\ln d_4-\ln
d_0));  \label{20}
\end{equation}

\begin{equation}
u(x,t)=-\frac{3(1-v^2)}{2a}\cosh ^{-2}(\frac 12(\beta (x-vt)-\ln d_5+\ln
d_0)).  \label{21}
\end{equation}
Especially $v=0$, the stationary solitary wave solutions of equation(5) are
obtained,

\begin{equation}
u(x,t)=-\frac{3(1-v^2)}{2a}\cosh ^{-2}(\frac 12(\beta x+\ln d_4-\ln d_0));
\label{22}
\end{equation}

\begin{equation}
u(x,t)=-\frac{3(1-v^2)}{2a}\cosh ^{-2}(\frac 12(\beta x-\ln d_5+\ln d_0)).
\label{23}
\end{equation}

Case 2, If $\beta =\sqrt{\frac{1-v^2}b}=0$, the solution of equation(15) is 
\begin{equation}
\omega (x,t)=s(\xi
)=d_0+d_1(x-vt)+d_2(x-vt)^2+d_3(x-vt)^3+d_4(x-vt)^4+d_5(x-vt)^5  \label{24}
\end{equation}
Substituting formula(24) into limited conditions(12)-(14), a set of algebra
equations are gotten, solving the set of equations, we get $d_2=0$, $d_3=0$, 
$d_4=0$, $d_5=0$. the solution of the set of equations(12)-(15) is 
\begin{equation}
\omega (x,t)=d_0+d_1(x-vt).  \label{25}
\end{equation}
Where $d_0$, $d_1$ are arbitrary constants. Putting (25) into (6), we get
the exact traveling solution of equation(5), 
\begin{equation}
u(x,t)=\frac{6bd_1^2}{a[d_0+d_1(x-vt)]^2}.  \label{26}
\end{equation}

Case 3, If $\beta =\sqrt{\frac{1-v^2}b}<0$, the solution of the set of
equations(12)-(15) is $\omega (x,t)=0$, only the trivial solution of
equation(5), $u(x,t)=0$, can be gotten.

Second, we discuss a new kind solution of equation(15). Now using $x$ as an
integrating factor, the equation(15) may be integrated once to yield

\begin{equation}
\omega _{xtt}+b\omega _{xxxxx}-\omega _{xxx}=p(t),  \label{27}
\end{equation}
where $p(t)$ is an arbitrary function. It's easy to know that the
equation(27) has the solution

\begin{equation}
\omega (x,t)=S(\xi )+q(t),  \label{28}
\end{equation}
where $S(\xi )$ is the traveling wave solution of equation $\omega
_{xtt}+b\omega _{xxxxx}-\omega _{xxx}=0$, $q(t)$ satisfies the equation $%
\frac{d^2}{dt^2}(q(t))=p(t)$. There exist some different cases to discuss
further.

Case 4, If $\beta =\sqrt{\frac{1-v^2}b}>0$, the solution of equation(15) is

\begin{equation}
\omega (x,t)=s(\xi )=d_0+d_1(x-vt)+d_2(x-vt)^2+d_3e^{\beta
(x-vt)}+d_4e^{-\beta (x-vt)}+q(t).  \label{29}
\end{equation}
Substituting formula(29) into limited conditions(12)-(14), a set of ordinary
differential equations are gotten, solving the set of equations, we get

\begin{eqnarray}
d_2 &=&0,  \label{30} \\
d_3 &=&\frac{bd_1^2(-1+4v^2)}{12d_4v^2(-1+v^2)},  \nonumber \\
q(t) &=&-\frac{d_1(-1+v^2)}vt.  \nonumber
\end{eqnarray}
Thus the solutions of equations(12)-(15) are gotten,

\begin{equation}
\omega (x,t)=d_0+d_1(x-vt)+\frac{bd_1^2(-1+4v^2)}{12d_4v^2(-1+v^2)}e^{\beta
(x-vt)}+d_4e^{-\beta (x-vt)}-\frac{d_1(-1+v^2)}vt,  \label{31}
\end{equation}
where $d_0$, $d_1$or $d_4$ are arbitrary constants. Substitute the solution
(31) into formula(6), the exact solution of equation(5) are obtained,

\begin{eqnarray}
&&u(x,t)=\frac{6b(d_1-d_4e^{-\beta (x-vt)}\beta +\frac{bd_1^2(-1+4v^2)\beta
e^{\beta (x-vt)}}{12d_4v^2(-1+v^2)})^2}{a(d_0+d_4e^{-\beta (x-vt)}-\frac{%
d_1(-1+v^2)t}v+\frac{bd_1^2(-1+4v^2)e^{\beta (x-vt)}}{12d_4v^2(-1+v^2)}%
+d_1(x-vt))^2}-  \label{32} \\
&&\frac{6b(\frac{d_4e^{-\beta (x-vt)}(1-v^2)}b+\frac{%
d_1^2(1-v^2)(-1+4v^2)e^{\beta (x-vt)}}{12d_4v^2(-1+v^2)})}{%
a(d_0+d_4e^{-\beta (x-vt)}-\frac{d_1(-1+v^2)t}v+\frac{bd_1^2(-1+4v^2)e^{%
\beta (x-vt)}}{12d_4v^2(-1+v^2)}+d_1(x-vt))}.  \nonumber
\end{eqnarray}

Case 5, If $\beta =\sqrt{\frac{1-v^2}b}=0$, the solution of equation(15) is

\begin{equation}
\omega (x,t)=s(\xi )=d_0+d_1(x-vt)+d_2(x-vt)^2+d_3(x-vt)^3+d_4(x-vt)^4+q(t).
\label{33}
\end{equation}
Substituting formula(33) into limited conditions(12)-(14), a set of ordinary
differential equations are gotten, solving the set of equations, we get $%
d_2=0$, $d_3=0$, $d_4=0$, $q(t)=2d_1t$. Thus the solutions of
equations(12)-(15) are gotten,

\begin{equation}
\omega (x,t)=s(\xi )=d_0+d_1(x-vt)+2d_1t,  \label{34}
\end{equation}
where $d_0$, $d_1$are arbitrary constants. Substitute the solution (31) into
formula(6), the exact solution of equation(5) are obtained,

\begin{equation}
u(x,t)=\frac{6bd_1^2}{a(d_0+2d_1t+d_1(x-vt))^2}.  \label{35}
\end{equation}

Case 6, If $\beta =\sqrt{\frac{1-v^2}b}<0$, the solution of equation(5) is
trivial also.

In summary, based the homogeneous balance method, we have introduced an
improved method: solving the original nonlinear PDE become solving the
linear PDE with some limited conditions. Obviously, the improved method can
be applied to the equations that the homogeneous balance method can be
applied to. For instance, we use it to the Bousseneq equation and derive
four kinds of exact solution. In case 1, the solutions are solitary wave
solution, using Wang's method can get them also; in case 2 and 5, the
solutions are traveling wave solutions; in case 4, the solution is new kind
solution to be discussed further; in case 3 and 6, the trivial solution is
gotten.

The authors are grateful to Professor Ming-Liang Wang for valuable
discussions. This work was supported by the National Natural Science
Foundation of China.

\end{document}